\documentclass[aps,prb,twocolumn,showpacs]{revtex4}
\usepackage{graphicx}% Include figure files

\begin{document}
\title{Collective   charge  excitations   along  cell   membranes }
\author{Efstratios Manousakis}
\affiliation{\\Department of Physics, Florida State University, 
Tallahassee, FL 32306-4350, and \\
Department of  Physics, University of Athens, Greece.}  
\date{\today}
\begin{abstract}
A significant part of the thin layers of counter-ions adjacent 
to the exterior and interior surfaces of a cell  membrane form  
quasi-two-dimensional (2D) layers of mobile charge.   
Collective  charge  density oscillations, known as  plasmon modes, in
these 2D charged systems of  counter-ions are predicted in the present
paper. This is based on  a calculation of the self-consistent response
of  this  system to  a fast  electric field  fluctuation. 
The possibility that the membrane channels might be using these 
excitations to carry out fast communication is suggested and    
experiments are proposed to reveal the existence of such
excitations.
\end{abstract}
\maketitle

Collective charge excitations
in condensed  matter, the so-called plasmon modes,  play a fundamental
role  in  understanding  the   properties  of  a  metal\cite{pines}. 
  These  are
excitations of the electron gas as a whole and are responsible for the
electromagnetic response of the  metal for frequencies above and below
the classical plasma oscillation  frequency, and for our understanding
of other  effects such  as screening.  The  plasmon is the  quantum of
such  collective charge oscillations  which may  exist in  any charged
system and in fact they  were first discovered in a classical 
plasma\cite{classical}.
In  quasi-two-dimensional systems  such  as surfaces  of metals
\cite{raether,geballe,ritchie,ferrell,stern2,fetter},  or
interfaces of metals with  insulators, plasmons have been investigated
both theoretically\cite{ritchie,ferrell,stern2,fetter} and 
experimentally\cite{raether,geballe}  
for several decades. 

Plasmon modes have been found\cite{helium} in a system of 
electrons on the surface of liquid helium.
The surface of a lipid bilayer with the surrounding  counter-ion system, 
which we consider in this paper, is analogous to that of 
electrons on helium surface as will be made clear below.
The surface of a cell membrane in aqueous environment becomes negatively 
charged\cite{hille-book}. 
It is well-known\cite{gouy,chapman,stern,debye-huckel,grahame} 
that a diffuse double layer of counter-ions
from the solution screens 
the negative charge, thus,  a thin two-dimensional (2D) layer of mobile cations
 is accumulated adjacent to the extra-cellular and intra-cellular membrane
surfaces. In addition, because of different permeabilities of the cell
membrane channels to various cations, the inside of the cell is kept 
at a negative potential relative to the outside.
In this paper it is  pointed out that  the  thin  layers  of  counter-ions 
adjacent to the  lipid bilayer are characterized  by  2D longitudinal 
plasmon excitations. These modes are excited 
in  response  to  a  fast perturbation such as the opening  of an ion 
channel.  The gating mechanism of ion channels along the cell membrane, 
that are believed to open using voltage sensing gates, is an open problem
in cell biology\cite{channel,hille-book}. 
The collective charge density wave which is theoretically
found in this paper can play the role of a very fast  signal carrier for such
ion channel communication.  Therefore the collective modes investigated 
here can play a fundamental role in cell biology.

\begin{figure}[ht] 
\begin{center}
\includegraphics[width=2. in]{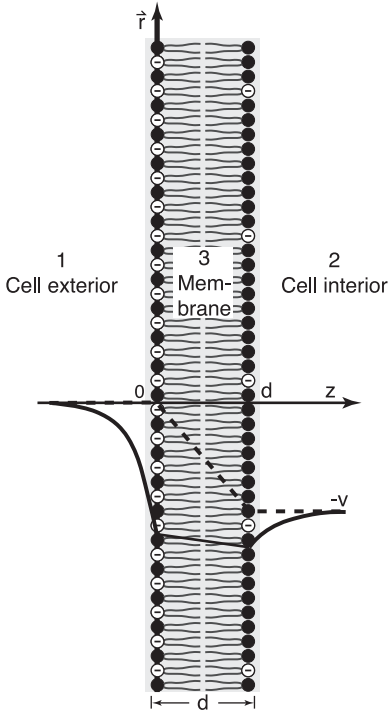}
 \caption{\label{figure1} 
The lipid bilayer  cell membrane  of width  d and  its various
features are explained in the  text.}
\end{center}
\end{figure}

The negatively charged exterior and
interior surfaces of biological membranes  of cells\cite{charge}  
as  shown   in  Fig.~\ref{figure1} where the density of charged
phospholipid heads  is higher on  the exterior membrane  surface.  
Diffuse double  layers of
counter-ions, such  as H$^+$,  Na$^+$, K$^+$, Ca$^{++}$,  ..., form on 
each membrane side.  Due  to the different membrane permeability  for 
these cations, there  is a voltage  $-V$ (of  the order  of $-100  mV$) 
between  the cell interior relative  to the exterior.  The  solid line 
in Fig.~\ref{figure1}  is the positive  free counter-ion distribution  
near the  membrane surfaces\cite{charge}.
In thermal equilibrium the counter-ion charge density  varies as
we move  away from the membrane  surface  as shown, within the
so-called Gouy-Chapman(GC)  length
$\lambda^i_{GC} =k_B T \epsilon_i /2\pi \sigma_i q_i$ for  each membrane 
side.  Here $\epsilon_i$ and $q_i$ are  
the dielectric constants and the counter-ion charges of the outside 
and inside of the cell. 
The surface  densities $\sigma_i=q_i n^0_i$ of both membrane sides 
are experimentally accessible and  
they can be as high as $n^0 \sim 10^{-2} \AA^{-2}$ or
even higher.   This counter-ion surface charge distribution is 
controlled by the bulk cation
concentrations of the intra-cellular and extra-cellular fluids\cite{grahame}.
In addition, depending on the temperature, cation concentration, and 
cation charge, 
a fraction of these ions can be adsorbed on the membrane surface.
The adsorbed ion density versus bulk aqueous ion concentration is determined
from the Stern adsorption isotherm theory\cite{stern} and its 
extensions\cite{cohen,tatulian}. 
The approximate value of $\lambda_{GC}$ is about $ 4 \AA$, 
if we use $T=300 K$, $\epsilon=80 $  and $n^0 =10^{-2}\AA^{-2}$. 
For charge fluctuations propagating with wave vectors ${\bf k}$
parallel  to the  surfaces and  magnitude $k<<1/\lambda_{GC}$ 
(which  is the  case  of our
interest),  these layers  can be regarded as two-dimensional.  

In this paper, the collective charge response of the mobile counter-ion charge 
of the interior and exterior surface layers to an  external electric
field perturbation is studied. It is shown that  there are longitudinal  
charge collective
modes with wave vectors parallel to the surface and the two layers are
coupled  and fluctuate  together as  a whole  in the  limit  where the
wavelength is much longer than the membrane thickness $d$. 

The counter-ions  of the  diffuse double
layer adjacent to the extra-cellular membrane surface 
are characterized  by  mass $m_1$  charge $q_1=Z_1 e$ ($Z_1$  
the  ionic valence and $e$ the electron charge)  and
instantaneous surface number density $n_1({\bf r},t)$ (${\bf r}=(x,y)$) 
and those adjacent to the intra-cellular membrane surface
by mass $m_2$, charge $q_2=Z_2 e$ and instantaneous surface 
number  density $n_2({\bf r},t)$. We are  interested in studying 
deviations $\delta n_i({\bf r},t)$ from the  
equilibrium surface density $n^0_i$, where 
$\delta n_i({\bf r},t) = n_i({\bf r},t)-n^0_i$.
Following the well-known approach\cite{pines} to this class of problems
where there is the possibility of collective response of the charge,
a perturbation $\phi_{ext}({\bf r}, t)$,
due to an external charge density fluctuation $\rho_{ext}({\bf r}, z,t)$
and current  fluctuations  ${\bf J}_{ext}({\bf r}, z,t)$,
is introduced. These physical quantities are related as follows
\begin{eqnarray}
\nabla^2 \phi_{ext}({\bf r},z,t) = - 4 \pi \rho_{ext}({\bf r},z,t)\\
\partial_t \rho_{ext}({\bf r},z,t) + {\bf \nabla} \cdot 
{\bf J}_{ext}({\bf r},z,t) = 0,
\end{eqnarray}
where the $z$ direction is taken perpendicular to the membrane surface.
This perturbation may be caused by the opening of a sodium or potassium 
membrane channel or by a real external probe. While these
channels are part of the cell membrane, they are not part of the degrees of
freedom considered here, namely we only consider the system of the
accumulated charge  on both sides of the cell membrane.

We would like to investigate the
response  of the  system to  such an  external perturbation  namely to
solve the  Poisson equation:
\begin{eqnarray}
\nabla^2 \phi_i({\bf r},z,t) = - {{4 \pi} \over {\epsilon_i}} 
\rho({\bf r},z,t),   
\label{field}
\end{eqnarray}
where $i=1,2$ and denotes the cell exterior and interior respectively.
Here $\phi_i({\bf r},z,t)$ is the total field in the 
regions $i=1,2$ and the total charge 
density is given by
\begin{eqnarray}
\rho({\bf r},z,t) = \rho_0(z) + \rho_{ext}({\bf r},z,t) + 
\rho_{ind}({\bf r},z,t) 
\end{eqnarray}
where $\rho_0(z)$ is the equilibrium charge density and 
$\rho_{ind}({\bf r},z,t)$ is the induced  charge density.   
Within our  2D  approximation, the
induced density  is written as 
\begin{eqnarray}
\rho_{ind}({\bf r},t) & = & q_1 \delta n_1({\bf r},t) \delta(z) +
q_2 \delta n_2({\bf r},t)\delta(z+d).
\label{ind}
\end{eqnarray}
 For both layers  we require the validity of the
linearized  continuity equation 
\begin{eqnarray}
\partial_t \delta n_i({\bf r},t) + n^0_i {\bf \nabla} \cdot  {\bf 
v}_i({\bf r},t) = 0, \hskip 0.5 in i=1,2
\label{cont}
\end{eqnarray}
where ${\bf v}_i({\bf r},t)$ is  the velocity  of the
counter-ions  along  the $i^{th}$ layer.  The  equation  of 
 motion for  a  single counter-ion maybe written as follows: 
\begin{eqnarray}
m_i \bigl (\partial_t {\bf v}_i + {{{\bf v}_i} \over {\tau}}\bigr )
= & - & q_i {\bf \nabla} \phi_i({\bf r},z=z_i,t) \nonumber \\
& - & {{m_i s_i^2} \over {n^0_i}} 
{\bf \nabla} \delta n_i({\bf r},t) \
\label{motion}
\end{eqnarray}
where $z_i=0^-,d^+$ for the exterior and
interior layer  respectively and $\tau$ is the time  between collisions, while
$s_i=\sqrt{ \gamma k_B T /m_i}$ is the velocity of density  fluctuations 
in the charged subsystems and 
the  value  of $\gamma = 2$  will  be  used  for  a 2D  ideal  gas.  
The first term is the force on the counter-ion due to the local electric
field while the second term is the force due to the local compression
of the counter-ion 2D gas. In  order  to
simultaneously  solve Eqs.  (\ref{field},\ref{cont},\ref{motion}),
 we consider  the Fourier  transform
(denoted by a tilde) of all  the functions of the 2D position vector
${\bf r}$ and time  $t$.  Eq.(\ref{field})  takes the form:
\begin{eqnarray}
\bigl ( \partial^2_z - k^2 \bigr ) {\tilde \phi}_i ({\bf k},z,\omega)  = 
-{{4 \pi} \over {\epsilon_i}} 
 {\tilde \rho}({\bf k},z,\omega) 
\label{v2}
\end{eqnarray}
where ${\bf k}$ is the  wave vector parallel to the membrane surface
and $\omega$ is the frequency. Eq. (\ref{v2})  has the following solution:
\begin{eqnarray}
 {\tilde \phi}_i ({\bf k},z,\omega)  & = & 
{{2 \pi} \over {\epsilon_i} k} \bigl 
[q_1 \delta \tilde n_1({\bf k},\omega) e^{-k|z|}
 + q_2  \delta \tilde n_2({\bf k},\omega) e^{-k|z-d|} \nonumber \\
      & + &
\int_{-\infty}^{\infty} 
dz'{\tilde \rho}'(z') e^{-k |z-z'|} \bigr ].
\label{field2}
\end{eqnarray}
where ${\tilde \rho}'(z)=\rho_0(z) + {\tilde \rho}_{ext}({\bf k},z,\omega)$.  
Substituting (\ref{field2})  in  the  Fourier transform  of  Eqs.
(\ref{cont},\ref{motion}) and by eliminating the  velocity variables we find
\begin{eqnarray}
\Bigl [ \omega (\omega-{i \over {\tau}}) & - & \omega^2_i(k) \Bigl ]
\delta {\tilde n}_i({\bf k},\omega) - \beta_i(k) 
\delta {\tilde n}_j({\bf k},\omega) \nonumber \\
& = &  \gamma_i({\bf k}, \omega),
\label{freq}
\end{eqnarray}
where $i,j=1,2$ or $2,1$. We  have used the definitions 
\begin{eqnarray}
\gamma_i({\bf k}, \omega) & = & {{Z_i} \over e} 
{\bar \omega}_i^2 \int dz' {\tilde \rho}'(z') e^{-k | z'- z_i|} \\
 \beta_i({\bf k}) & = & Z_1 Z_2 {\bar \omega}^2_i e^{-k d},\\
\omega_i^2(k) & = & Z_i^2 {\bar \omega}_i^2 + s_i^2 k^2,\\
{\bar \omega}_i^2 & = & {{2\pi e^2 n^0_i k} \over {m_i \epsilon_i}}.
\end{eqnarray}
The  system of   Eqs. (\ref{freq}) has  a solution  for the
responses  $\delta \tilde n_i({\bf k}, \omega)$ which  attain  their   
maximum  values  for  the  following
frequencies:
\begin{eqnarray}
\Omega_{\pm}(k) & = & 
\sqrt{\omega^2_{\pm}(k) + \Bigl ({1 \over {2\tau}}\Bigr )^2}
+ {i \over {2\tau}} 
\label{freq1}
\\
\omega^2_{\pm}(k) & = & {{\omega^2_1(k) + \omega^2_2(k)} \over 2}
\nonumber \\
 & \pm &  \sqrt{\Bigl ({{\omega^2_1(k) - \omega^2_2(k)} \over 2} \Bigr )^2 +
 \beta_1(k) \beta_2(k)}  .
\label{freq2}
\end{eqnarray}
The frequencies $\omega_i(k)$  are the  frequencies of
plasmon oscillations of each layer when their mutual coupling 
(terms $\beta_i(k)$) is small, i.e., when $kd >>1$.

\begin{figure}[ht] 
\begin{center}
\includegraphics[width=3.0 in]{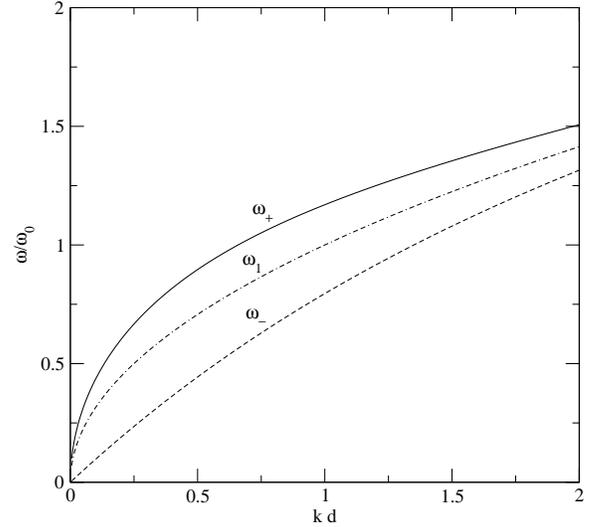}
 \caption{\label{figure2}  
Eigen-frequencies   of   the   membrane   collective   charge
oscillations  when all  characteristic parameters  of the  carriers of
both layers are the same. The curve in the middle denoted by $\omega_1$ is the
plasmon frequency of each layer in the limit where the coupling between the
inner and outer layers is neglected. The frequencies denoted by $\omega_-$
($\omega_+$) correspond to in-phase (out-of-phase) oscillations of 
both layers together. When the parameters of the two layers are different,
the corresponding plot looks similar, where instead of having one 
$\omega_1$ curve in the middle we have two, one for $\omega_1$ and 
a second for $\omega_2$.}
\end{center}
\end{figure}

 The frequencies are shown in Fig.~\ref{figure2} for the
case  where all  the parameters  of both  layers are  taken to  be the
same. The  frequencies are  given in units  of 
\begin{eqnarray}
\omega_0 = Z \sqrt{ {2 \pi e^2 n^0} \over {md\epsilon}}
\end{eqnarray}
and the  wave-number in
units of $1/d$. In terms of these variables the frequency of each independent
layer  can be  expressed as 
\begin{eqnarray}
{{\omega^2(k)}\over {\omega^2_0}} = kd (1 + kd_0) \\
d_0 \equiv k_B T \epsilon /(2 \pi^2 e^2 n^0) \simeq 5 \AA
\end{eqnarray} 
and the   second term  has a  small
contribution for  the range of $k$ in  the graph because $d \simeq 30 \AA$.
 The frequencies of
each  independent layer  are  shown  by the  middle  curve. The  lower
frequency  corresponds  to in-phase  charge  oscillations  of the  two
layers   and    the   higher   frequency    to   out-of-phase   charge
oscillations.   It   is   straightforward   to  show   that   in   the
long-wavelength  limit  ($kd <<1$) 
\begin{eqnarray}
\omega_-(k \to 0) = c_p k \\
\omega_+(k \to 0) = \sqrt{2 s_0^2 k/d + (s_1^2 -s_0^2) k^2},
\end{eqnarray}
and in  the  particular  case  where  the
parameters  characterizing   the  layers   are  equal,  we   have  
$c_p = \sqrt{s_0^2 + s_1^2}$ with $s_0 = \omega_0 d$. 
Taking realistic  values  of  the parameters  we  find that  for H$^+$
counter-ions the  frequency-scale is $\omega_0/2 \pi \sim 10^{11} Hz$ 
and  the speed is $c_p \sim 3000 m/s$.  For heavier counter-ions these
scales will be smaller as both scale as $1/\sqrt{m}$.  
\begin{figure}[ht] 
\begin{center}
\includegraphics[width=3.0 in]{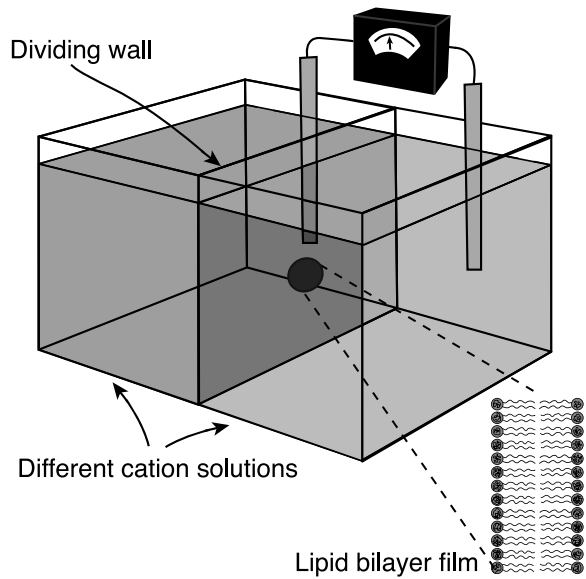}
 \caption{\label{figure3}  
An idealized experimental set up to control the cation concentration on both
membrane sides. }
\end{center}
\end{figure}

The  group and
phase velocity  of the  higher frequency mode  and of  the independent
layer oscillations  in the long-wavelength  limit diverge as 
$u(k \to 0) = s_0 /\sqrt{k d}$.  Therefore in the long-wavelength limit 
these velocities  are higher  than  the counter-ion  thermal velocities  and
therefore the  collision frequency is  lower than the  collective mode
frequency.  Therefore  the  Landau   damping  rate  is  small  in  the
long-wavelength limit.  In order to obtain a  more accurate expression
of  the plasmon  frequencies  instead  of the  bare  mass and  surface
density, one should use the effective mass of the hydrated ion and the
effective nearly-free counter-ion charge density. If there are binding
sites  on the  membrane surface  this will  reduce the  effective free
surface charge density.  Therefore more realistically we should expect
frequencies below the above mentioned values. In addition, the surface
corrugations of the electrostatic interaction of the counter-ions with
the phospholipid  surface introduce a  non-zero frequency gap $\Omega_0$  
at zero plasmon momentum.  This $\Omega_0$ is the  frequency of local  
charge oscillation
near  a  local  minimum  of  the  interaction  potential  between  the
counter-ion and the membrane  surface. When the surface charge density
is  low, this  interaction may  lead to  the formation  of  a Wigner
crystal   where  the   counter-ions   will  order   in  a   triangular
or other lattice structures\cite{blagoev}. 
These structures of ordered charged layers have been observed in 
electronic bilayer systems, consisting of two quasi-two-dimensional 
layers of electron or hole liquids\cite{mitchell}.
Such a Wigner crystal  will be characterized by plasmon modes
with frequency  given by  the same  form as Eqs.  (\ref{freq1},\ref{freq2}) with  
the masses
replaced by the  ion effective mass within the  Wigner crystal.  

These collective modes  can be  investigated experimentally by  studying the
electromagnetic absorption spectrum in  the frequency region of 
$10^{10}-10^{11}Hz$
(depending  on  the  wavelength of  the  excited plasmon and the 
counter-ion parameters and distribution)  of
artificial  phospholipid  bilayer  membranes  which  separate  aqueous
solutions  of  different ionic  concentrations (see Fig.~\ref{figure3})  
with high  counter-ion
concentrations\cite{hille-book} $\sigma_i$ near  the exterior and 
interior  surfaces. Up to $10mm^2$ size membrane can be formed in the hole
of the wall separating the two aqueous compartments shown in 
Fig.~\ref{figure3} by appropriately
mixing the phospholipids as was first discovered by 
Mueller et al.\cite{mixing}. An appropriate voltage can be
applied between the two aqueous compartments of different cation 
solutions separated by the membrane. Plasmon modes of the two thin 
counter-ion layers adjacent to the membrane can be excited and probed 
by a source/spectrometer system.
With the experimental techniques and source/spectrometer 
systems available today, it is possible to study the AC electromagnetic
response of very small samples over a range of frequencies and
to cover the above frequency domain with one or a combination of
source/spectrometer systems.  The plasmon
frequencies  should scale  as $\sqrt{\sigma}$ with  respect to  
surface concentrations
(which  can be  estimated from  the bulk  concentrations) and 
 as $1/\sqrt{m_i^*}$ with
respect  to various  counter-ion effective  masses.  

These  2D plasmon
oscillations can be used by the  various parts of the cell membrane to
carry  out communication.  For  example, these  modes  are excited  in
response to a cell membrane  pore opening which creates a local charge
distribution fluctuation. The phenomenological studies of the behavior
of the gates  of the voltage-sensitive ion channels  in cell membranes
indicate a high sensitivity\cite{hille-book} of the voltage sensor 
to small changes of
the  membrane  electric  field.   This  implies  that  the  2D  plasma
oscillations could  play a role in  triggering the sensor  to open the
gate.    Before,  however,   one  proceeds   further  to   study  such
possibilities it  is important to  carry out an experimental  study to
verify the  existence and the  properties of these plasmon  modes.

\end{document}